\documentclass[aps,prl,twocolumn,showpacs,superscriptaddress,repreint]{revtex4-1}

\usepackage{graphicx}
\usepackage{dcolumn}
\usepackage{bm}
\usepackage{amsmath}
\usepackage{hyperref}
\usepackage{color}

\renewcommand{\vec}[1]{{{\bf #1}}}
\makeatother
\usepackage{tikz}

\begin{document}

\title{Active particles with soft and curved walls: Equation of state,
  ratchets, and instabilities}

\author{Nikolai Nikola}
\affiliation{Department of Physics, Technion, Haifa, 32000, Israel}
\author{Alexandre P. Solon}
\affiliation{Universit\'e Paris Diderot, Sorbonne Paris Cit\'e, MSC, UMR 7057 CNRS, 75205 Paris, France}
\affiliation{Massachusetts Institute of Technology, Department of Physics, Cambridge, Massachusetts 02139, USA}

\author{Yariv Kafri}
\affiliation{Department of Physics, Technion, Haifa, 32000, Israel}

\author{Mehran Kardar}
\affiliation{Massachusetts Institute of Technology, Department of Physics, Cambridge, Massachusetts 02139, USA}

\author{Julien Tailleur}
\affiliation{Universit\'e Paris Diderot, Sorbonne Paris Cit\'e, MSC, UMR 7057 CNRS, 75205 Paris, France}

\author{Rapha\"el Voituriez}
\affiliation{Laboratoire de Physique Th\'eorique de la Mati\`ere Condens\'ee,
  UMR 7600 CNRS /UPMC, 4 Place Jussieu, 75255 Paris Cedex, France}
\affiliation{Laboratoire Jean Perrin, UMR 8237 CNRS /UPMC,
  4 Place Jussieu, 75255 Paris Cedex}

\date{\today}

\begin{abstract}
  We study, from first principles, the pressure exerted by an active
  fluid of spherical particles on general boundaries in two
  dimensions. We show that, despite the non-uniform pressure along
  curved walls, an equation of state is recovered upon a proper
  spatial averaging. This holds even in the presence of pairwise
  interactions between particles or when asymmetric walls induce
  ratchet currents, which are accompanied by spontaneous shear
  stresses on the walls. For flexible obstacles, the pressure
  inhomogeneities lead to a modulational instability as well as to the
  spontaneous motion of short semi-flexible filaments. Finally, we
  relate the force exerted on objects immersed in active baths to the
  particle flux they generate around them.
\end{abstract}


\maketitle

Active forces have recently attracted much interest in many different
contexts \cite{Marchetti2013RMP}.  In biology, they play crucial roles
on scales ranging from the microscopic, where they control cell shape
and motion~\cite{Julicher:2007rx} to the macroscopic, where they play
a dominant role in tissue
dynamics~\cite{Poujade2007PNAS,Heisenberg:kq}.  More generally, active
systems offer novel engineering perspectives, beyond those of 
equilibrium systems. In particular, boundaries have been shown to be
efficient tools for manipulating active particles. Examples range from
rectification of bacterial
densities~\cite{Galajda2007,Tailleur2009EPL} and optimal delivery of
passive cargoes~\cite{Koumakis2014SM} to powering of microscopic
gears~\cite{DiLeonardo2010PNAS,Sokolov2010PNAS}. Further progress,
however, requires a predictive theoretical framework which is
currently lacking for active systems. To this end, simple settings
have been at the core of recent active matter research.

Understanding the effect of boundaries on active matter starts with
the mechanical pressure exerted by an active fluid on its containing
vessel. This question has been recently studied extensively for dry
systems~\cite{Mallory2014,Yang2014,Takatori2014,Solon2015a,SwimPressure,Yan2015JFM,winkler2015virial,speck:2015},
revealing a surprisingly complex physics. For generic active fluids,
the mechanical pressure is {\it not} a state
variable~\cite{Solon2015a}.  The lack of equation of state, through a
dependence on the wall details, questions the role of the mechanical
pressure in any possible thermodynamic description of active
systems~\cite{Takatori2015,Solon_interactions}; it also leads to a
richer phenomenology than in passive systems by allowing more general
mechanical interplays between fluids and their containers.

Interestingly, for the canonical model of self-propelled spheres of
constant propelling forces, on which neither walls nor other particles
exert torques, the pressure acting on a {\it solid flat} wall has been
shown to admit an equation of
state~\cite{Yang2014,Takatori2014,Solon_interactions}.  While the
physics of this model does not clearly differ from other active
systems, showing for instance wall accumulation~\cite{Elgeti2013EPL}
and motility-induced phase
separation~\cite{Fily2012PRL,Tailleur2008PRL,Cates2015MIPS}, the
mechanical pressure exerted on a flat wall satisfies an equation of
state, even in the presence of pairwise
interactions~\cite{Yang2014,Takatori2014,Solon_interactions}. One
might thus hope that the intuition built on the rheology of
equilibrium fluids extends to this case.  Derived in a particular
setting, the robustness of this equation of state however remains an
open question. For instance, the physics of active fluids near curved
and flat boundaries is very
different~\cite{Mallory2014,Fily2014,Fily2015SM,Yan2015JFM,SwimPressure}.
Specifically, particles accumulate non-evenly depending on the
curvature of confining walls, generating a spatially varying
pressure~\cite{Fily2014,Fily2015SM}. Furthermore, the interplay
between active particles and flexible objects, such as polymers and
membranes, shows a rich non-equilibrium
phenomenology~\cite{PolymerSwelling,PolymerBrush,PolymerCollapse,PolymerCrowdedDynamics,PolymerLooping,kaiser:2015,ActivatingMembranes,Kikuchi24112009,Mallory2015PRE}.
Characterising the role and properties of active forces in these
contexts is thus an open and challenging question.

In this letter we study, from first principles, the confinement of
torque-free active particles beyond the case of solid flat walls. We
first show analytically that, while the pressure on curved walls is
inhomogeneous, one recovers an equation of state for the average force
{\it normal} to the wall, even in the presence of pairwise
interactions.  This surprising result also holds for asymmetric walls
which act as ratchets and, as we show, generate currents and forces
tangential to the wall. Contrary to the average normal force, these
shear stresses depend on the details of the potential used to model
the wall and therefore do not admit an equation of state.  Moreover,
we show that the pressure inhomogeneities trigger interesting new
physics.  For flexible partitions, we show how a finite-wavelength
modulational instability sets in, followed by a long-time
coarsening. Interestingly, this also explains the atypical folding and
self-propulsion of semi-flexible filaments immersed in active
baths~\cite{PolymerCollapse,PolymerLooping}. Finally, we give a simple
relation between the force exerted on an asymmetric object in an
active bath and the current of active particles it generates around
it.

We start by considering non-interacting active particles, of
positions $\vec{r}_i=(x_i,y_i)$ and headings
$\vec{e}_{\theta_i}=(\cos\theta_i,\sin\theta_i)$, which follow
the Langevin equations
\begin{eqnarray}\label{eq:Lengevin_Non_Inter}
  \dot{\vec{r}}_i &=&v\vec{e}_{\theta_i}-\mu_{t}\vec{\nabla}V+\sqrt{2D_{t}}\vec{\eta}_i(t)\\
  \dot{\theta}_i &=&\sqrt{2D_{r}}\eta^{r}_i(t)
\end{eqnarray}
in addition to randomly changing orientation (tumbling) with rate
$\alpha$. Here $v$ is the propulsion speed, $\mu_{t}$ is the
translational mobility~\footnote{In all the numerics, $\mu_t=1$.},
$D_t$ and $D_r$ are the translational and rotational diffusivities,
$V(\vec{r})$ is a static potential which defines the confining walls,
and the $\eta$'s are unit-variance Gaussian white noises. This model
encompasses the well-studied run-and-tumble (RTP) dynamics and active
Brownian particles (ABP), with pure rotational diffusion. The
persistence length of the particle, or run length, is given by
$v/(D_{r}+\alpha)$.

We first consider a system with periodic boundary conditions along the
$\hat{y}$ direction and structured walls along the ${\hat{x}}$
direction. The wall potential starts, say, along the right edge of the
system, at $x_{w}(y) = x_{0}+A\sin(2\pi y/L_{p})$, and takes the form
$V(\vec{r})=\frac{1}{2}\lambda [x-x_{w}(y)]^{2}$ for $x>x_w$, with a
mirrored opposing wall at $-x_{w}(y)$. The system height in the
$\hat{y}$ direction is taken to be an integer times $L_{p}$.
Equations of state, in or out of equilibrium, only exist in the
thermodynamic limit and we always take the distance between the walls
much larger than any correlation length. The bulk is then uniform,
isotropic, and independent of what happens in the boundary layers
close to the walls; we thus consider only one edge of the system.

\begin{figure}
  \includegraphics[width=.49\columnwidth,height=.35\columnwidth]{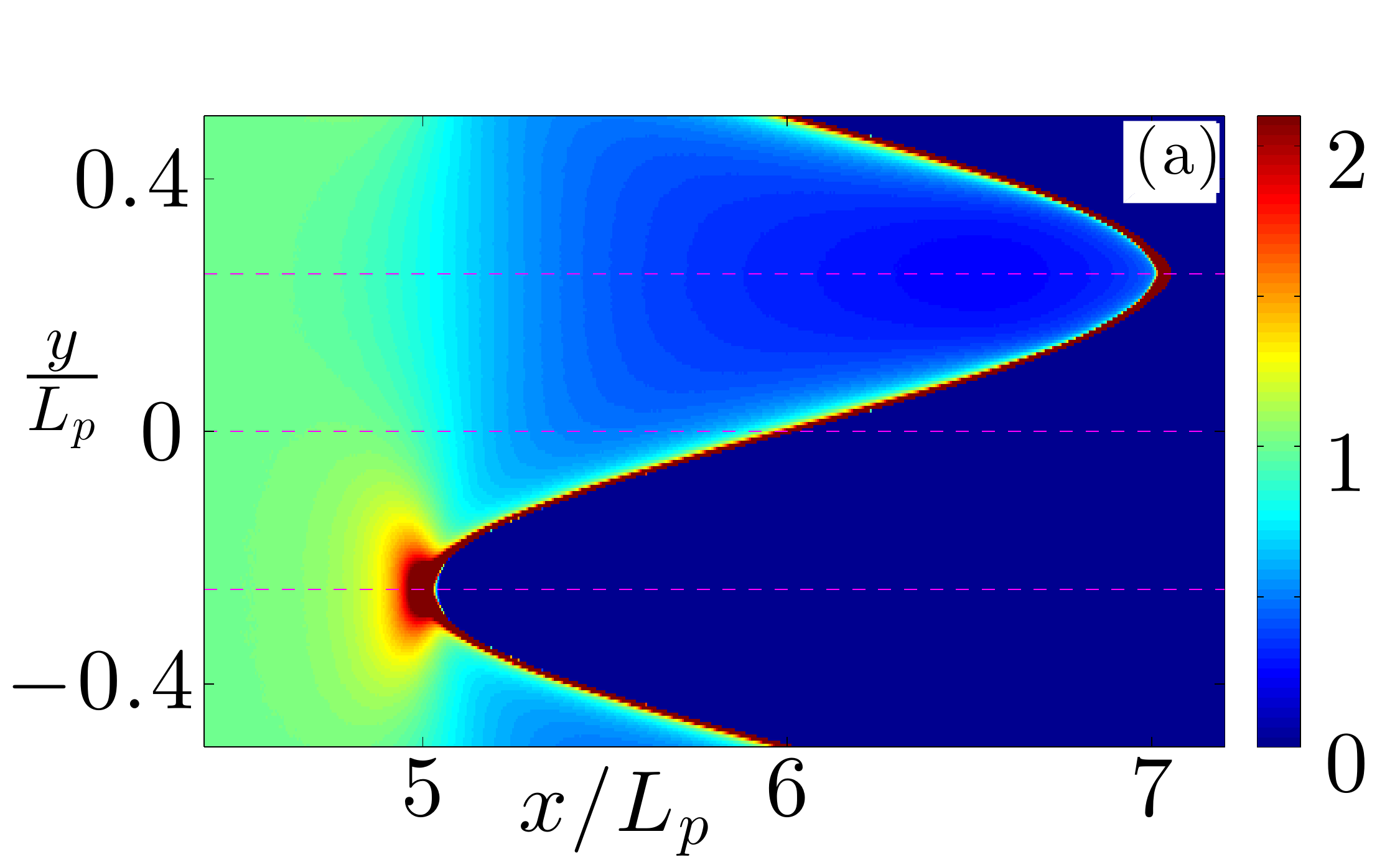}
  \includegraphics[width=.49\columnwidth,height=.34\columnwidth]{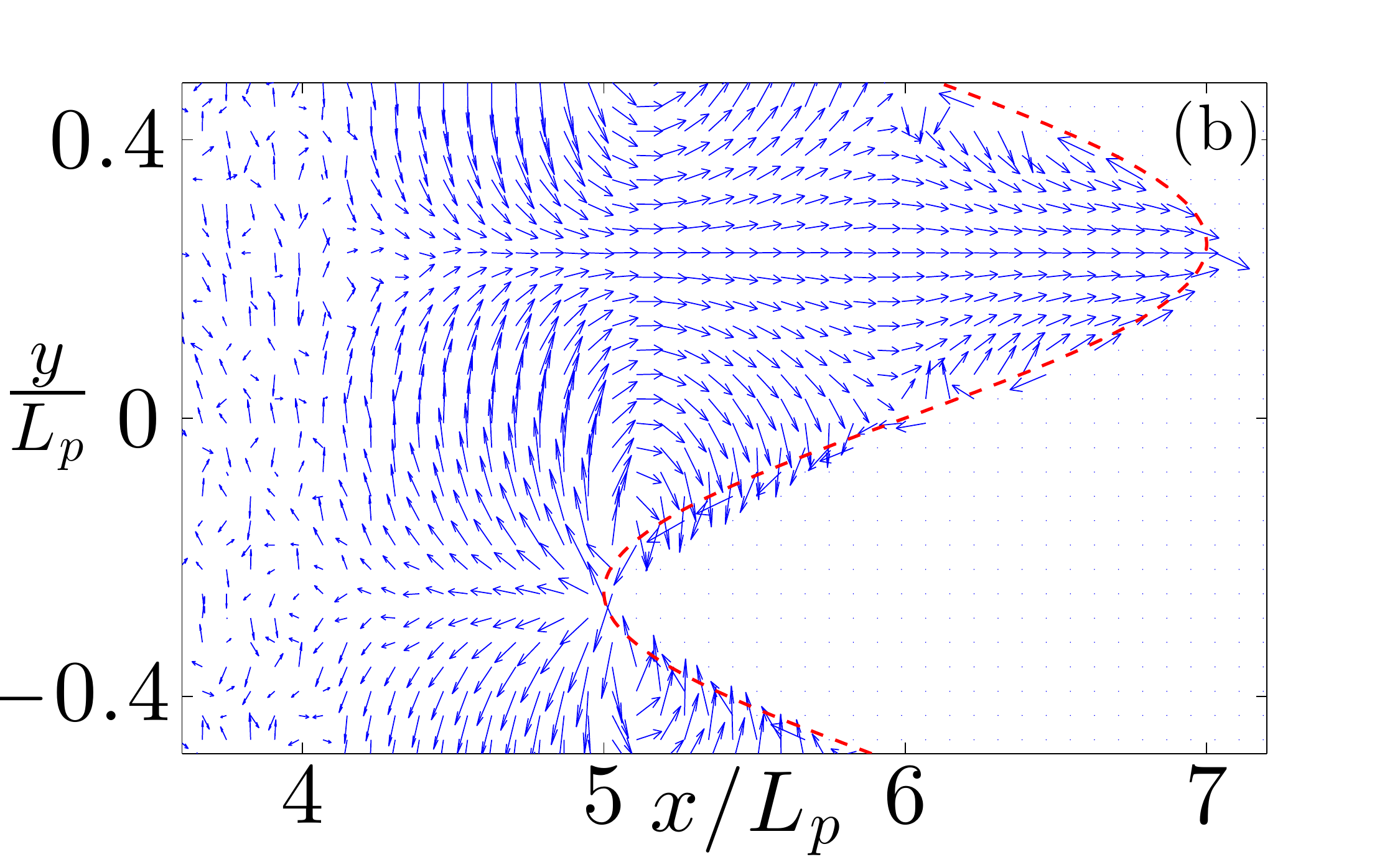}\\
  \includegraphics[width=.49\columnwidth,height=.35\columnwidth]{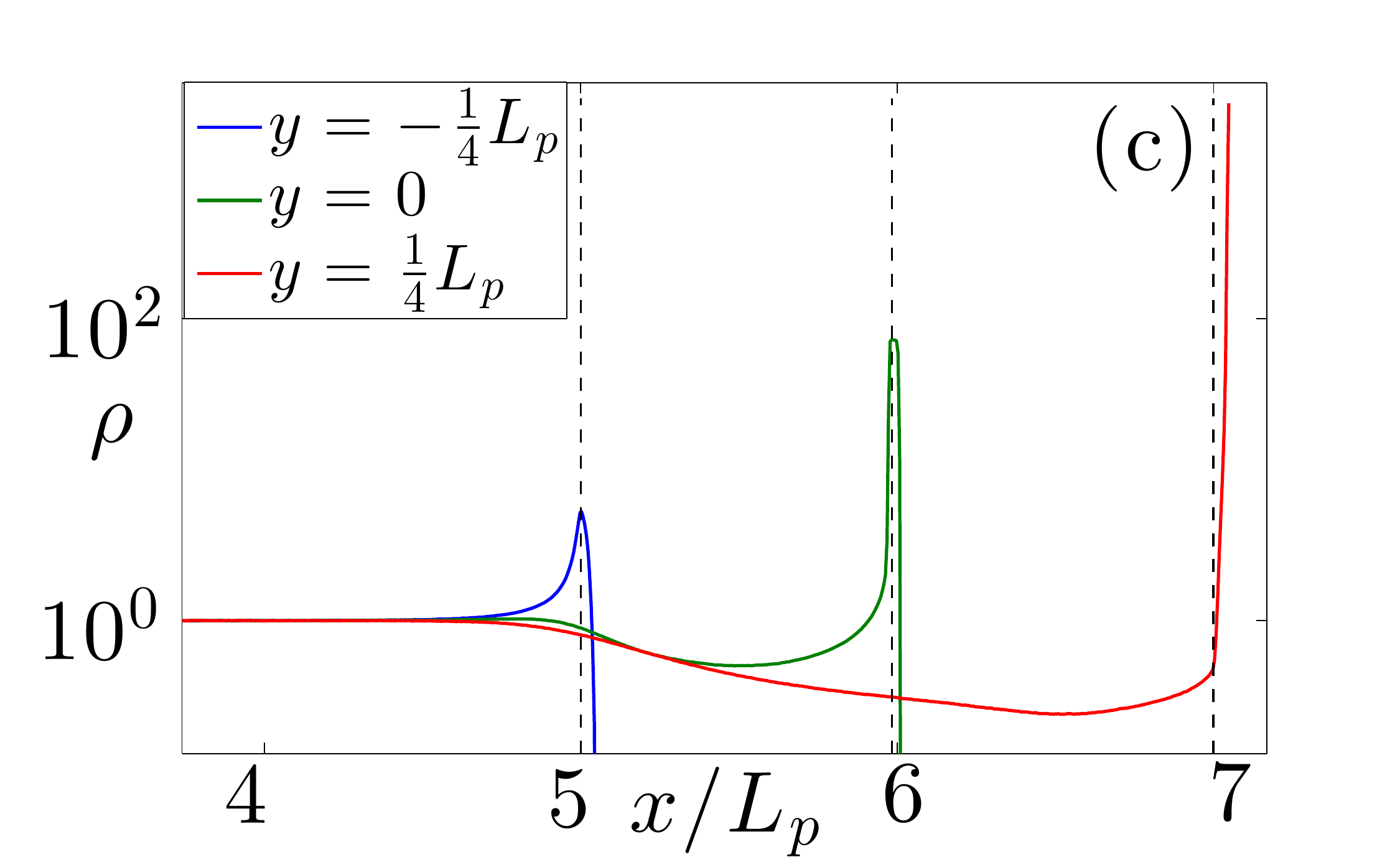}
  \includegraphics[width=.49\columnwidth,height=.35\columnwidth]{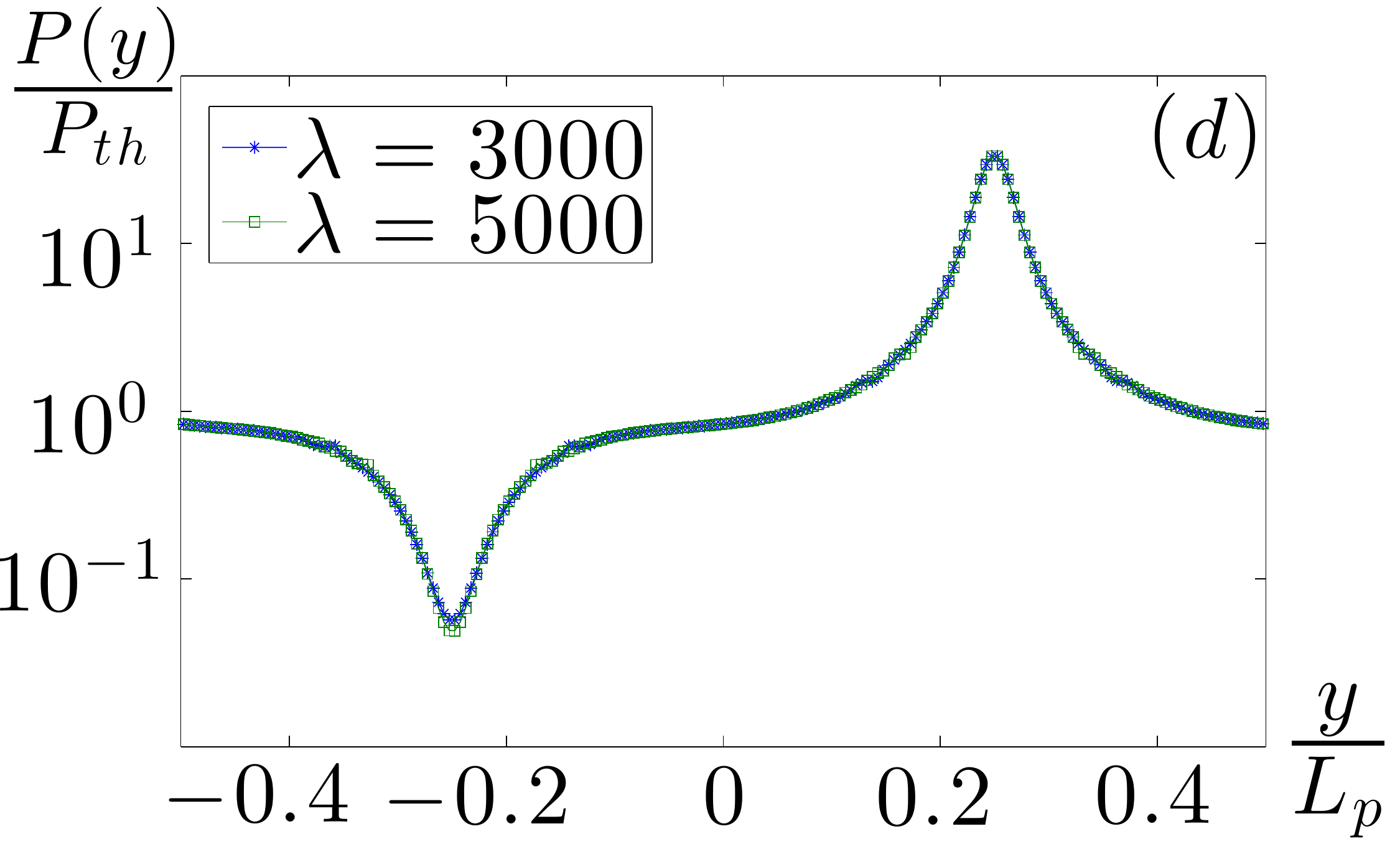}\\
  \caption{Density {\bf (a)} and current {\bf(b)} of non-interacting ABPs near
    the right edge of the system with the hard wall potential
    described in the text with $v=D_r=24$, $D_t=0$, $L_p=0.5$,
    $A=0.5$, $\lambda=1000$. The red dashed curve correspond to
    $x_w(y)$. {\bf(c)} Three cross sections of the particle
    density taken at the three horizontal dashed lines in (a). The
    vertical lines correspond to $x_w(y)$. {\bf(d)}
    Pressure normal to the wall, normalized by
    Eq.~\eqref{eq:PressureTeff}, as a function of $y$, in the hard
    wall regime.}
  \label{fig:Density}\label{current} \label{density_cross}\label{P_y_profile}\label{fig:Densities}
\end{figure}

\begin{figure}
  \raisebox{.5cm}{\includegraphics[width=.2\columnwidth,trim=700 0 0 0]{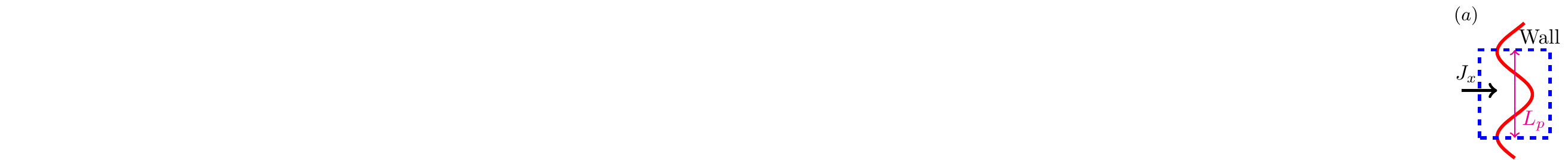}}
  \includegraphics[width=.38\columnwidth,totalheight=.35\columnwidth]{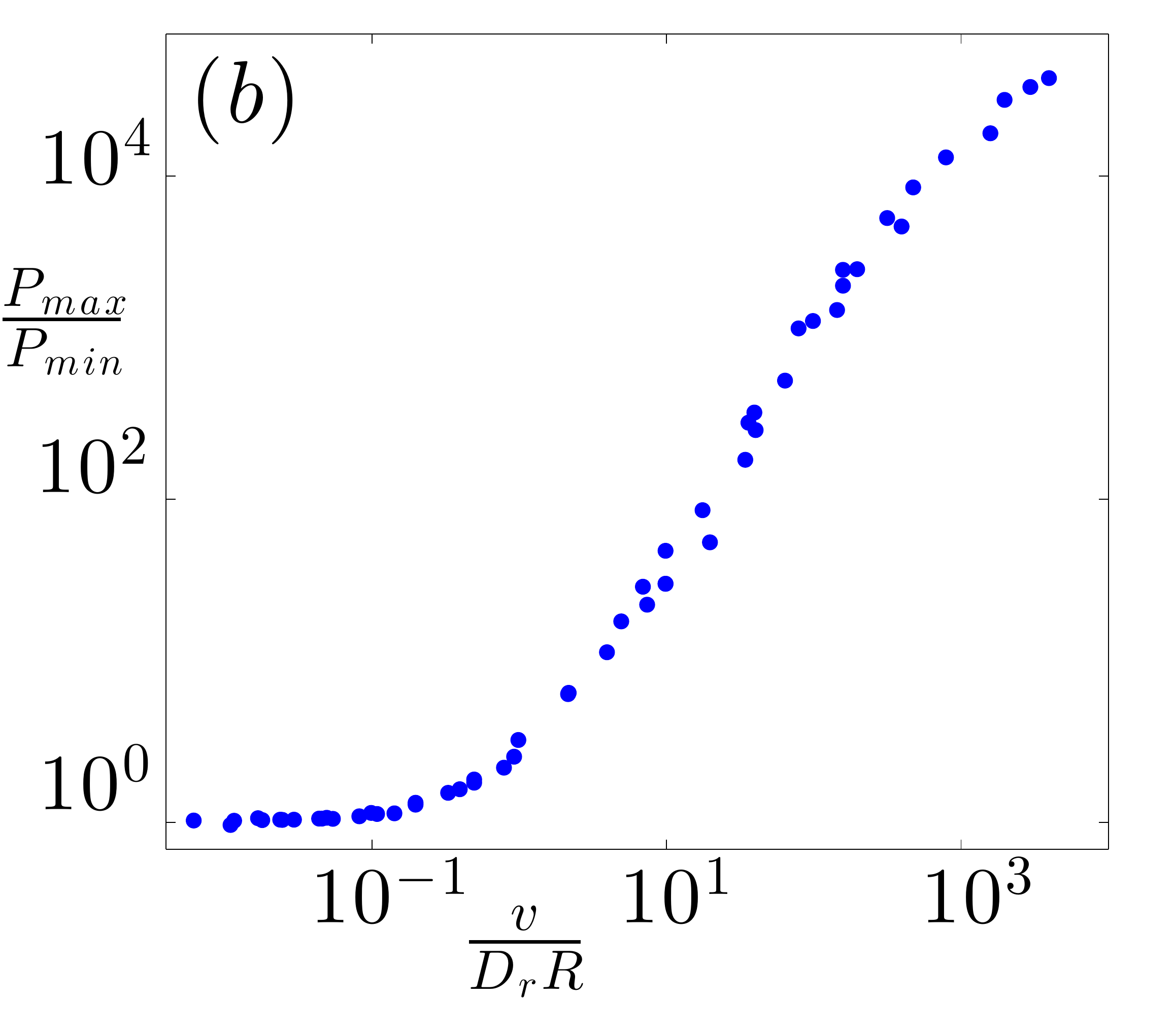}
  \includegraphics[width=.38\columnwidth,totalheight=.35\columnwidth]{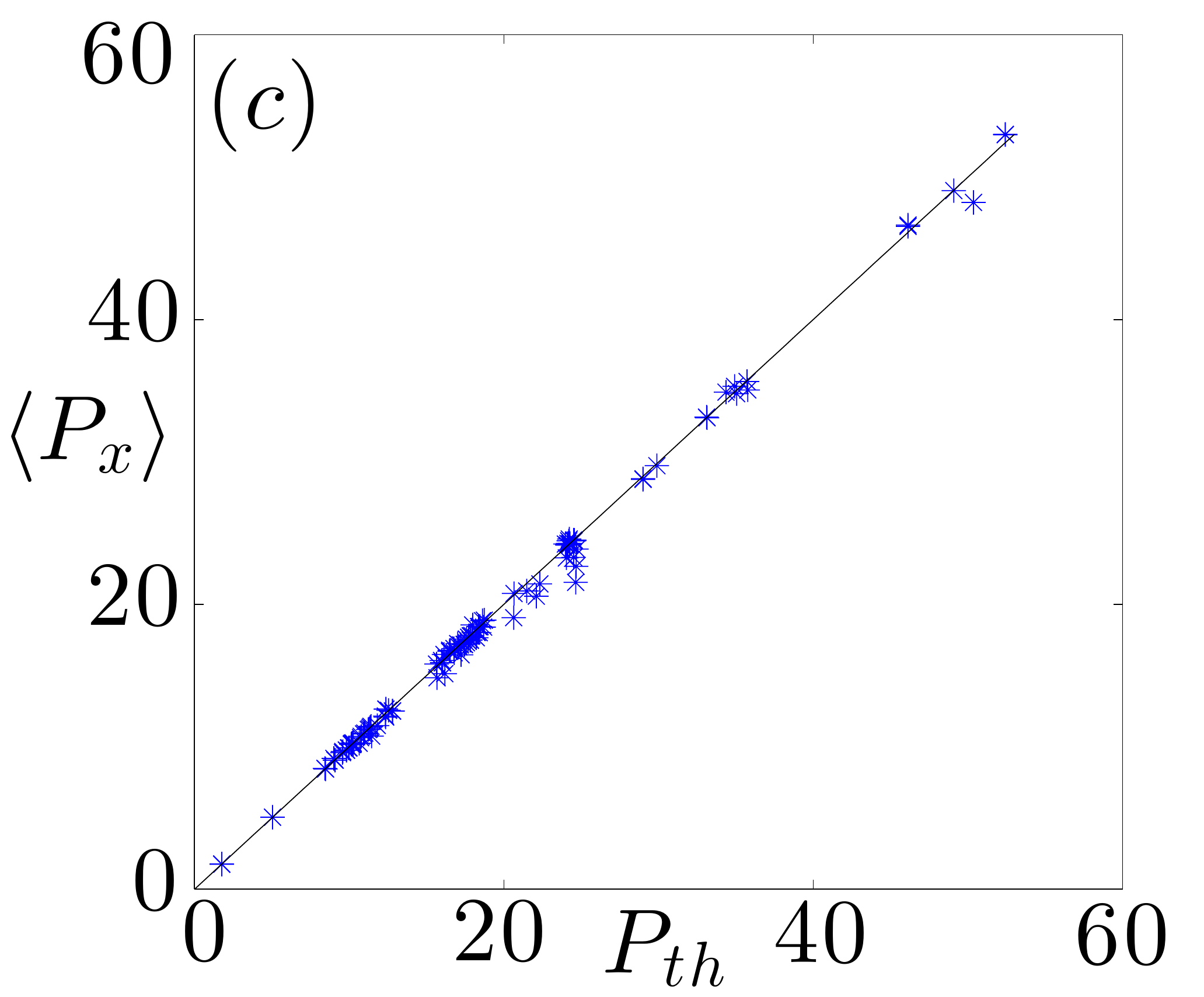}
  \caption{\label{fig:Scaling_and_P_Pth} {\bf (a)}: For a
    periodic wall, the probability flux is periodic along $y$ and
    vanishes in the wall so that $\int_0^{L_p} J_x dy=0$ in
    steady-state.  {\bf (b)}: The ratio between the maximal and
    minimal values of $P_x(y)$ on hard sinusoidal walls as a function
    of $v/D_r R$, where $R=L_p^2/(4\pi^2 A)$ is the radius of
    curvature, measured for a variety of combinations with
    $v\in[0.05,40]$, $A\in [0.3,12]$, $D_{r}\in [0.1,20]$, and
    $L_{p}\in[0.3,80]$, and $\lambda=1000-3000$. {\bf (c)}: Mean
    pressure versus its theoretical prediction
    Eq.~\eqref{eq:PressureTeff}. Data collected from 70 simulations of
    ABPs, RTPs, passive Brownian particles, and combinations thereof,
    with symmetric and asymmetric, hard and soft, periodic walls.}
  \label{fig:Scaling}
  \label{Px_Pth}
\end{figure}

Let us first consider \emph{hard} walls, with $\lambda$ large enough
that particles are arrested by the wall potential on a scale orders of
magnitude smaller than any other relevant length scales.  Examples of
numerically measured steady-state particle and current densities are
shown in Figs.~\ref{fig:Density}.a and~\ref{fig:Density}.b. As
expected~\cite{SwimPressure,Fily2014,Fily2015SM}, the density along
and close to the wall is non-uniform and in general unequal at points
of equal potential (as opposed to what happens at thermal
equilibrium). Remarkably, in addition to the thin layer close to the
wall where particles accumulate, complex potential-dependent
steady-state densities and currents are found in the whole wall
region. There is a depletion of particles in the \textit{outer
  concave} region of the wall, where particles stream towards the
outer apices, and a density increase close to the \textit{inner
  convex} apices, due to the recirculation of particles along the
walls. Most importantly, the local pressure varies considerably along
the wall~\footnote{This holds for non-interacting active
  particles with \textit{macroscopic} wall modulations; its origin
  lies in the activity of the particles and has nothing to do with
  the non-uniformity of pressure seen at the microscopic scale for
  equilibrium interacting particle systems, see [J.-P. Hansen and
  I. R. McDonald. {\it Theory of simple liquids.} Elsevier, 1990]}  (see
Fig.~\ref{P_y_profile}.d). For hard walls, the force is always normal
to the wall surface and the local pressure can be evaluated as
$P(y)=\int_{{\bf r}_\star}^\infty \rho(\vec{r}) \nabla V \cdot d{\bf
  r}$
where the integral is taken in the direction normal to the wall,
${\bf r}_\star$ is inside the bulk of the system, and $\rho(\vec{r})$
is the density of particles. Somewhat counter-intuitively, the
pressure is highest close to the depleted region, at the outer apices,
because the depletion is compensated by a stronger accumulation at the
wall, where the potential is non-zero (see
Fig.~\ref{density_cross}.c). Similarly, it is lowest at regions where
there is an accumulation of particles near the wall. Finally, we find,
numerically, that the ratio between the maximal and minimal pressures
(at the outer and inner apices, respectively) is a function of the
dimensionless parameter $\frac{v}{D_{r}R}$ (Fig.~\ref{fig:Scaling}b),
where $R=L_p^2/(4\pi^2 A)$ is the radius of curvature at the apices.

Naively, these results suggest that the equation of state obtained
in~\cite{Solon2015a} for this model is valid only for flat
walls. However, Fig.~\ref{Px_Pth} shows that the pressure, despite its
non-uniformity, satisfies the same equation of state as in the case of
flat walls~\cite{Solon2015a} once averaged over a period of the
wall. More precisely, the force per unit length acting on a period of
the potential, defined as
\begin{equation}
  \langle P_x \rangle\!=\!\frac{1}{L_{p}}\!\int_{0}^{L_{p}}\!\!\!P_{x}(y)dy\;, \; P_{x}(y)=\int_{x_\star}^\infty \!\!\!\rho({\bf r}) \partial_x V dx
  \label{eq:PressureDef}
\end{equation}
with $P_{x}(y)$ the force per unit length exerted by active particles
on the wall along the $\hat{x}$ direction \footnote{$P_x(\vec r)$ is
  the component of the normal force per unit area in the
  $x$-direction, which for simplicity we denote as ``pressure''.},
obeys an equation of state. (By symmetry, for the potentials
considered so far, the mean force along the $\hat{y}$ direction is
zero). The averaged pressure is thus independent of the wall
potential, whether hard or soft, a result which persists for any
$V(\vec{r})$ despite the fact that the local pressure depends on the
exact form of the wall potential~\cite{supp}.  It would be
interesting to see if this can be related to the applicability of
the virial theorem to the systems we consider
here~\cite{Yang2014,Takatori2014,winkler2015virial,speck:2015,falasco:2015}.

To prove this result, consider the dynamics of the probability density
to find a particle at $\vec{r}$ with
orientation $\theta$:
\begin{equation}
  \begin{aligned}
    \partial_{t}\mathcal{P}(\vec{r},\theta) &= -\vec{\nabla}\cdot (v \vec{e}_{\theta}-\mu_{t}\vec{\nabla}V-D_{t}\vec{\nabla})\mathcal{P}(\vec{r},\theta) \\
    &+(D_{r}\partial_{\theta}^{2}-\alpha)\mathcal{P}(\vec{r},\theta) +\frac{\alpha}{2\pi}\int_{0}^{2\pi}\!\!\!\!{d\theta^{\prime}\mathcal{P}(\vec{r},\theta^{\prime})}.
  \end{aligned}
  \label{eq:FPE}
\end{equation}
Integrating over $\theta$ yields in
steady-state $\nabla\cdot \vec J=0$, where
\begin{equation}
  \vec J(\vec r)=v \int_{0}^{2\pi}\!\!\!\vec{e}_{\theta}\mathcal{P}d\theta-\mu_{t}\tilde \rho\nabla V-D_{t}\nabla\tilde\rho.
  \label{eq:current}
\end{equation}
and $\tilde \rho(\vec r)=\int d \theta \mathcal{P}(\vec
r,\theta)$. For flat walls, the invariance along $\hat y$ of the
system imposes $J_x=0$, which directly leads to an equation of state
for the local force per unit length in the ${\hat x}$
direction~\cite{Solon2015a}. While $J_x$ can be locally non-zero for
structured walls, the mean flux of particles through a closed path
still has to vanish in steady state. It is then always possible to
find a length $L_p$ such that $\int_0^{L_p} dy J_x(x,y)=0$, where
$L_p$ can for instance be the period of a periodic potential or the
full wall length (see Fig.~\ref{fig:Scaling_and_P_Pth}a). Following~\cite{Solon2015a}, one can then construct an equation
of state for $\langle P_x \rangle$ instead of the local pressure
$P_x$. For instance, for non-interacting particles
\begin{equation}
  \langle P_x \rangle=\rho_{0}\Big [\frac{v^{2}}{2\mu_{t}(D_{r}+\alpha)}+\frac{D_{t}}{\mu_{t}}\Big]\label{eq:PressureTeff}\equiv P_{th}
\end{equation}
prefectly fits our simulations (see Fig.~\ref{Px_Pth}),
where $\rho_0$ is the mean number density of particles in
the bulk. The full proof, allowing for pairwise interactions, is
given in~\cite{supp}.

\begin{figure}[t]
  \label{fig:asym_contour}\includegraphics[width=.46\columnwidth,height=.35\columnwidth]{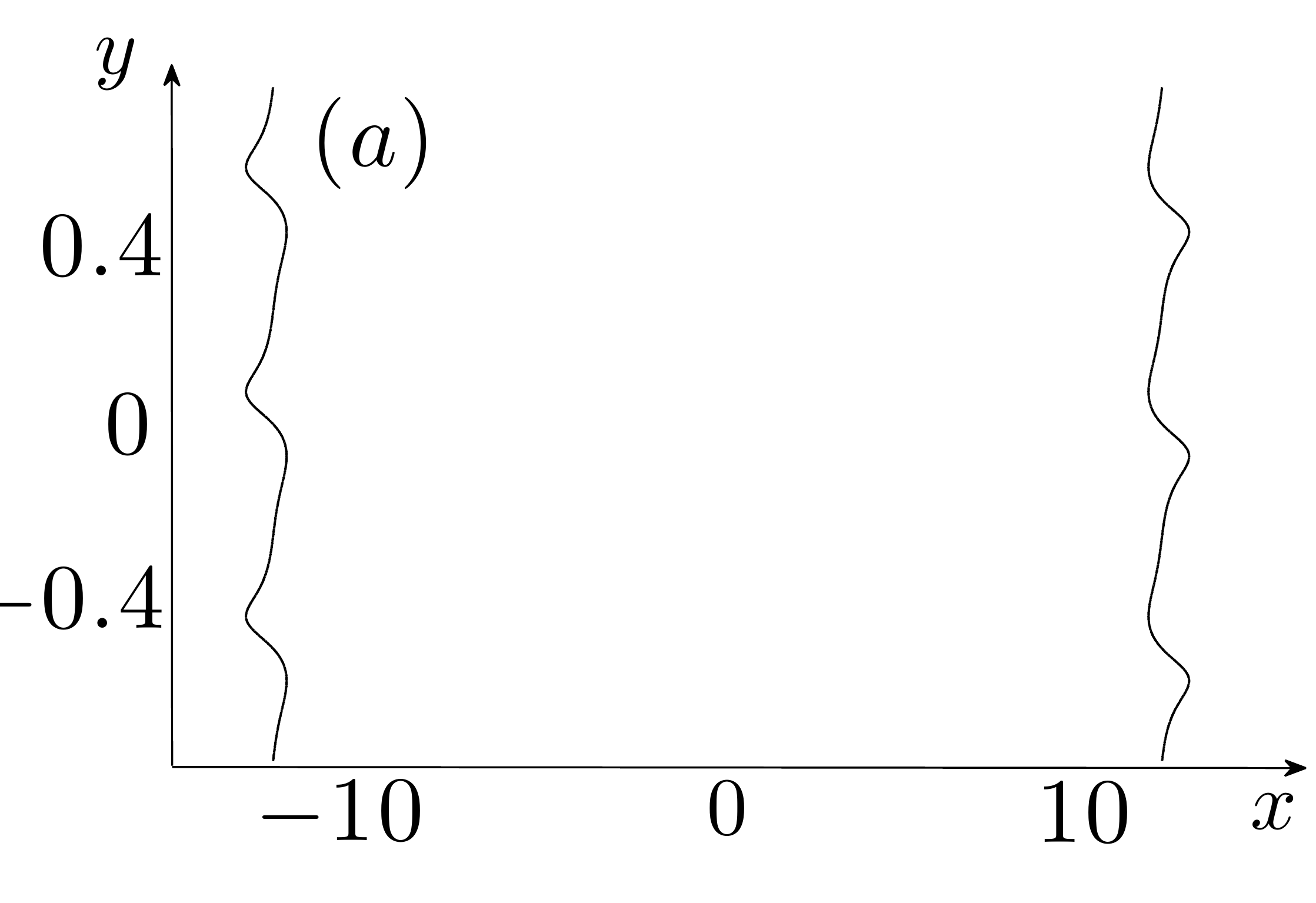}
  \label{PandJy}\includegraphics[width=.52\columnwidth,height=.35\columnwidth]{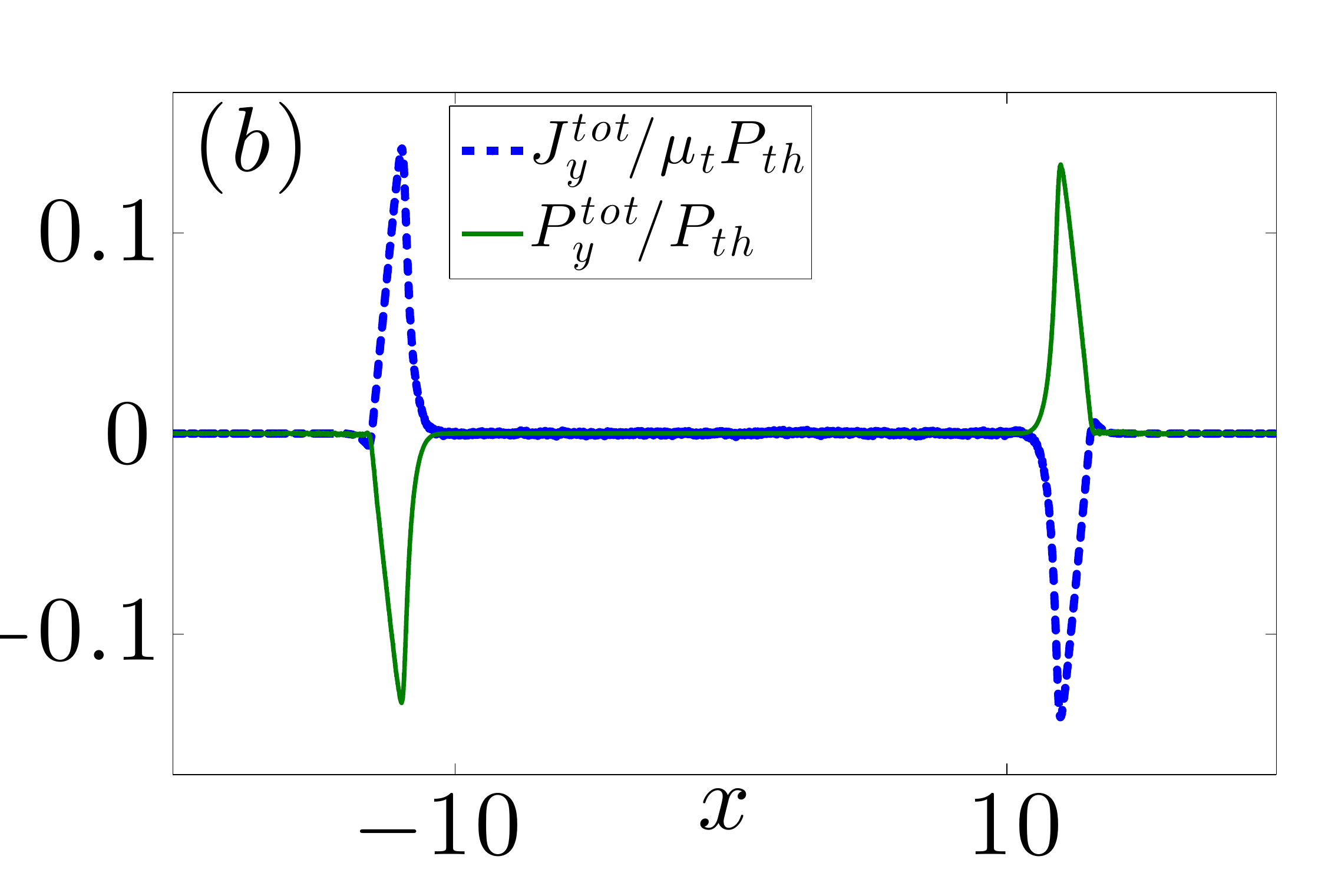}
  \caption{{\bf (a)}: Equipotential line ($V=5$) of the
    asymmetric potential~\eqref{eq:AsymmetricWall}. {\bf (b)}: Ratchet current  and shear stress as functions of $x$. Eq.~(\ref{eqn:forcey}) is verified numerically within 1\%. Simulation of ABPs with $x_0=10$, $\lambda=2, A=0.3, L_p=0.5$,  $v=D_{r}=24$.
    \label{fig:Parallel_pressure_current}}
\end{figure}

Interestingly, the equation of state is valid for any wall, including
asymmetric ones which, in the spirit of
ratchets~\cite{Galajda2007,Ratchet_reich,Ratchet_transport,Angelani2011EPL,Yariv:2014},
may induce a net particle current along the wall. Such a current,
whose direction can be controlled by the asymmetry of the wall, is
accompanied by a shear force exerted by the active particles on the
wall, parallel to its general surface. This is illustrated in
Fig.~\ref{fig:Parallel_pressure_current} for an asymmetric wall
potential given by
\begin{equation}
  V(x,y)=\frac{1}{2}\lambda (x-x_0)^{2}\Big[1+Ae^{\cos\big(\frac{2\pi y}{L_{p}}\big)}\sin\Big(\frac{2\pi y}{L_{p}}\Big)\Big]\label{eq:AsymmetricWall}.
\end{equation}
This spontaneous shear stress exerted by an active fluid on an
asymmetric surface is impossible in equilibrium systems. It can
be quantified as (see Fig.~\ref{fig:Parallel_pressure_current})
\begin{equation}
  P_y^{\rm tot}(x)=\int_0^{L_p}dy \rho(\vec
  r) \partial_y V(\vec r).
\end{equation}
This explains the spontaneous rotation of microscopic
gears~\cite{DiLeonardo2010PNAS,Sokolov2010PNAS} and relates to
the ratchet current through:
\begin{equation}\label{eqn:forcey}
  \int_{x_\star}^\infty dx P_y^{tot}= -\frac 1 {\mu_t} \int_{x_\star}^\infty dx  J_y^{tot}.
\end{equation}
Here $J_{y}^{tot}(x)=\int_0^{L_p}J_y(\vec r)dy$ is the total current
in the $\hat y$ direction (see~\cite{supp} and
Fig.~\ref{fig:Parallel_pressure_current}) and Eq.~\eqref{eqn:forcey}
is the total force exerted by the active fluid on one period of the
wall in the $\hat y$ direction. Note that $\langle P_x \rangle$,
despite all these complications, still satisfies the same equation of
state as for symmetric walls.

As shown above, although the averaged pressure satisfies an equation
of state, the details of the interaction with the wall, e.g. the local
pressure, are quite unlike an equilibrium system. Specifically, we
note that for hard surfaces, the highest pressure is always at the
concave apex of the wall surface, while the lowest pressure is at the
convex apex. We now show that for flexible objects this leads to a
generic modulational instability.

\begin{figure}[t]
  \includegraphics[width=.66\columnwidth]{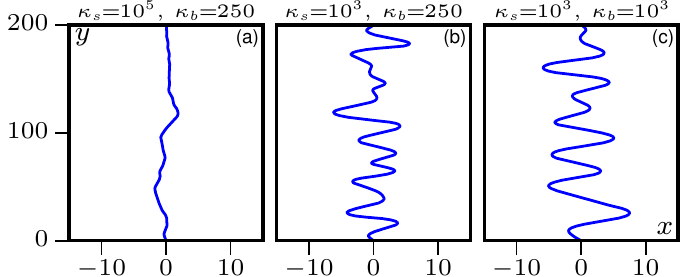}
  \includegraphics[width=.31\columnwidth]{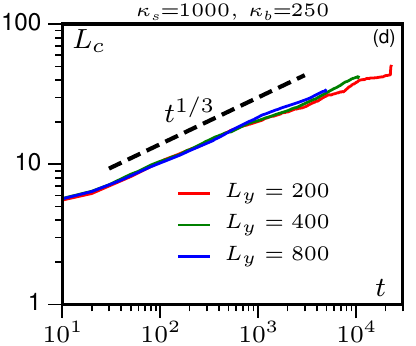}
  \caption{Snapshots of a flexible interface in an active bath, not
    shown for clarity. The stretching constant controls the threshold
    {\bf (a,b)} while the bending constant controls the wavelength of the
    instability {\bf (b,c)}. Snapshots taken at $t=30$ after starting from a straight
    filament.
    %
    The characteristic length $L_c$, defined as the first zero
    of the auto-correlation function of the transverse monomer
    displacement, slowly coarsens as time goes on (d). Parameters:
    $\rho_0=1$, $v=10$, $D_r=1$, $r_1=r_0=0.3$, $\kappa_s=1000$,
    $\kappa_b=250$.}\label{fig:coarsening}\label{flexible_wall_inst}
\end{figure}

The origin of the instability can be understood by considering a
flexible interface (a filament in 2D), characterized by stretching and
bending rigidities, and anchored at the top and bottom of a container
holding the active particles (See Fig.~\eqref{flexible_wall_inst}).
The results above imply that, once induced by a fluctuation, a local
deformation will have a finite pressure difference on the two sides of
its apices, which will tend to further increase the
deformation. Within linear stability analysis, the fate of a
fluctuation of wavenumber $q$ can be understood as follows: The
unstabilizing contribution of the active pressure scales as the
curvature of the interface (see Fig.~\ref{fig:Scaling}
and~\cite{supp}), hence as $q^2$. It competes with the stabilizing
effects of the tension and the bending rigidities, which scale as
$q^2$ and $q^4$, respectively. This implies that, for large enough
activity, the interface is unstable below a certain wavenumber $q_c$
with a fastest growing mode $q_{\rm max}$ controled by the interplay
between the tension, activity and the bending
rigidity. (See~\cite{supp} for details and estimates of $q_c$.)

To observe the instability, we carried out numerical simulations
(videos provided in~\cite{supp}), in which a semi-flexible filament with fixed
ends is immersed in an active gas of ABPs. The filament is modeled as
a chain of beads whose potential energy is given by
\begin{equation}
  E=\sum_{i=1}^{N-1} \frac {\kappa_s}{r_1} \frac{(|{\bf r}_{i+1}-{\bf r}_{i}|-r_0)^2}2-\frac {\kappa_b}{r_1} ({\bf t}_{i+1}\cdot {\bf t}_{i})
\end{equation}
where ${\bf r}_i$ is the position of bead $i$, $r_0$ the rest length
of the springs, $r_1$ the initial distance between the beads
($r_1>r_0$ for a chain initially under tension) and ${\bf t}_i$ is the
unit vector tangent to the $i^{\rm th}$ bond. The beads interact with
the active particles via a stiff repulsive harmonic potential which
prevents the active particles from crossing the flexible chain. As the
simple argument above suggests, for a given system size, at large
stretching constant $\kappa_s$, the wall undergoes small fluctuations
around its mean position. As $\kappa_s$ is decreased, the pressure
imbalance around the apices of the filament fluctuations is not
compensated anymore and initial microscopic fluctuations evolve into
larger scale, wave-like features whose initial wavelength is
controlled by the bending constant $\kappa_b$ (see
Fig. \ref{flexible_wall_inst}). The initial instability is then
observed to slowly coarsen, with an exponent compatible with a $1/3$
power-law (Fig.~\ref{fig:coarsening}). Coarsening in active system is
notoriously difficult to
assess~\cite{Thompson2011JSM,Redner2013PRL,Stenhammar2013PRL} and a
more precise characterization of this phenomenon will be addressed in
future works.

It is interesting to consider what the instability predicts
for flexible filaments which are not pinned at their
extremities. Typical snapshots for increasing filament lengths $L_f$
are shown in Fig.~\ref{fig:polymer} and in supplementary
movies. First, the bending rigidity prevents significant modulations
of very short filaments, allowing only for slow diffusive motion. As $L_f$ increases beyond the smallest
unstable wavelength, the filaments bend, leading to the spontaneous
formation of a wedge. The pressure difference on both sides of the
filament then propels it forward. This explains the propulsion of
``parachute-shaped'' filaments observed numerically in
Ref.~\cite{PolymerLooping}. As the size of the polymer increases
further the parachute shape becomes unstable: a full period of an
unstable mode develops and one observes short-lived spontaneous
rotors. Finally, upon increasing $L_f$ beyond the period of the
fastest growing mode, the pressure imbalance folds the polymer. This
instability thus partly explains the atypical folding of polymers in
active baths reported numerically in the
literature~\cite{PolymerLooping,PolymerSwelling,PolymerBrush,PolymerCollapse,PolymerCrowdedDynamics,PolymerLooping}.
This transition as the size of the filament increases can be monitored
in the diffusivity of its center of mass, which exhibits a sharp peak
corresponding to self-propelled wedges (Fig.~\ref{fig:polymer}).

Remarkably, our formalism allows us to relate the forces exerted on
asymmetric objects, such as the parachute-shaped filaments, to the net
flow of active particles around them, a result that goes far beyond
the sole cases explored in this paper. To see this, integrate
Eq.~(\ref{eq:current}) over a surface containing an isolated
object. This leads to
\begin{equation}\label{eq:JF}
  {\vec F}^{\rm tot}= -\boldsymbol{\cal J}/\mu_t\,,
\end{equation}
where $\vec F^{\rm tot}\equiv \int d^2\vec r \rho\nabla V$ is the
total force exerted on the object and $\boldsymbol{\cal J}\equiv\int
d^2\vec r \vec J(\bf r)$ the total current of active
particles~\cite{supp}. In the limit of slow, quasistatic
  motion of the object, Eq.~\eqref{eq:JF} relates in a simple formula
  the ability of an asymmetric object to act as a ratchet to its
  self-propulsion by an active bath.

\begin{figure}
  \includegraphics[width=.45\columnwidth]{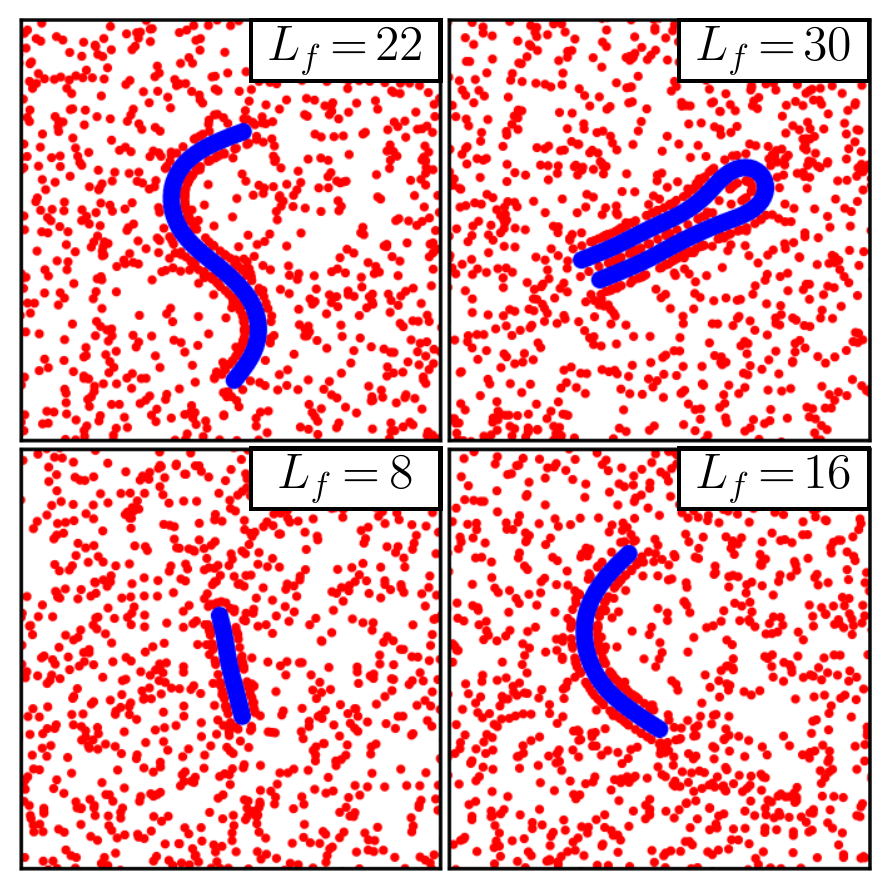}\hspace{0.3cm}
  \raisebox{0.15cm}{\includegraphics[width=.35\columnwidth]{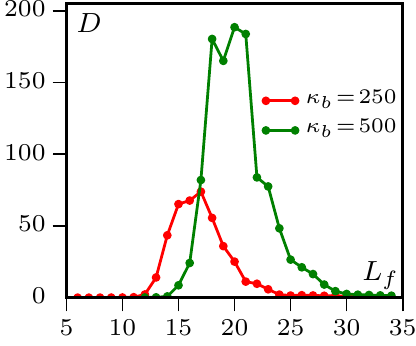}}
  \caption{{\bf Left}: Typical configurations for filaments of varying
    length for $\kappa_b=250$, $v=10$, $D_r=1$, $\rho_0=1$,
    $\kappa_s=1000$, $r_1=r_0=0.3$ and box size $50\times 50$ with periodic
    boundary conditions. {\bf Right}: Diffusivity of the filament as a function of its length.}\label{fig:polymer}
\end{figure}

\textit{Conclusions}--- To summarize, we have shown, both analytically
and numerically, that while the forces exerted by an active fluid on a
structured wall are in general inhomogeneous, an equation of state is
recovered upon a proper spatial averaging. This result holds for
non-interacting active particles as well as in the presence of
pairwise interactions. Walls lacking an `up-down' symmetry act as
ratchets and generate transverse fluxes. While the mean force normal
to the wall axis still satisfies an equation of state, there is now a
wall-dependent shear stress. The numerics shown in this letter for
ABPs are complemented in~\cite{supp} by similar results for RTPs which
highlight their generality. For flexible boundaries, we have shown how
the fluctuations of the wall shape can be enhanced by pressure
inhomogeneities which trigger a modulational instability. For freely
moving objects and filaments, this instability sheds new light on a
host of phenomena which have been observed numerically, such as the
atypical looping and swelling of polymers in active
baths~\cite{PolymerLooping,PolymerSwelling,PolymerBrush,PolymerCollapse,PolymerCrowdedDynamics,PolymerLooping}
as well as predicts new behaviors.

This work paves the way to new interesting questions in the
engineering and control of active fluids. It would be interesting, for
example, to determine how the shear stress generated by walls with
asymmetric roughness can be optimized, or whether the dependence of
the active motility of semi-flexible filaments on their size can be
used as a sorting mechanism.

\emph{Acknowledgements.} JT was supported by ANR project Bactterns. NN
and YK are supported by an I-CORE Program of the Planning and
Budgeting Committee of the Israel Science Foundation and an Israel
Science Foundation grant. APS acknowledges funding through a PLS
fellowship from the Gordon and Betty Moore foundation. MK is supported
by NSF grant DMR-12-06323. 

\bibliography{letter_bib}

\end{document}